\documentclass{article}

\usepackage{packages}

\DeclareMathOperator*{\enc}{\textsf{Enc}}
\DeclareMathOperator*{\dec}{\textsf{Dec}}
\DeclareMathOperator*{\aux}{aux}

\newcommand{\B}{\mathbb{B}}                                     % Binaries
\newcommand{\shield}[2]{\mathcal{S}_{#1, #2}}

\newcommand{\gta}{\text{GTA}}
\newcommand{\publicD}{\Delta}
\newcommand{\RandMech}{\mathcal{M}}
\newcommand{\randMech}[2]{\mathcal{M}_{#1, #2}}
\newcommand{\offset}{\omega}

\makeatletter
\newcommand\incircbin
{%
	\mathpalette\@incircbin
}
\newcommand\@incircbin[2]
{%
	\mathbin%
	{%
		\ooalign{\hidewidth$#1#2$\hidewidth\crcr$#1\bigcirc$}%
	}%
}
\newcommand{\olor}{\incircbin{\lor}}
\makeatother

\newtheorem{definition}{Definition}

\newtheorem{proposition}{Proposition}

\definecolor{marsala}{HTML}{964F4C}

\begin{document}

\title{When approximate design for fast homomorphic computation provides differential privacy guarantees}

%On an imperfect yet fast homomorphic argmax operator providing differential privacy guarantees
%When approximations of a lightweight homomorphic argmax operator provide differential privacy guarantees
%When lightweight design for fast homomorphic computation provides differential privacy guarantees
%Thank God, our vote counter makes mistakes, we can vote secretly !

\author{
    Arnaud Grivet S\'{e}bert$^{1}$\and
    Martin Zuber$^{1}$\and
    Oana Stan$^{1}$\and
    Renaud Sirdey$^{1}$\and
    C\'{e}dric Gouy-Pailler$^{1}$\and
    \\
    $^1$Institut LIST, CEA, Université Paris-Saclay,\\F-91120, Palaiseau, France
    \\
    \{arnaud.grivetsebert, martin.zuber, oana.stan, \\
    renaud.sirdey, cedric.gouy-pailler\}@cea.fr
}

\maketitle

\begin{abstract}
While machine learning has become pervasive in as diversified fields as industry, healthcare, social networks, privacy concerns regarding the training data have gained a critical importance. In settings where several parties wish to collaboratively train a common model without jeopardizing their sensitive data, the need for a private training protocol is particularly stringent and implies to protect the data against both the model's end-users and the actors of the training phase. Differential privacy (DP) and cryptographic primitives are complementary popular countermeasures against privacy attacks. Among these cryptographic primitives, fully homomorphic encryption (FHE) offers ciphertext malleability at the cost of time-consuming operations in the homomorphic domain. In this paper, we design SHIELD, a probabilistic approximation algorithm for the argmax operator which is both fast when homomorphically executed and whose inaccuracy is used as a feature to ensure DP guarantees. Even if SHIELD could have other applications, we here focus on one setting and seamlessly integrate it in the SPEED collaborative training framework from \cite{grivet2021speed} to improve its computational efficiency. After thoroughly describing the FHE implementation of our algorithm and its DP analysis, we present experimental results. To the best of our knowledge, it is the first work in which relaxing the accuracy of an homomorphic calculation is constructively usable as a degree of freedom to achieve better FHE performances.
\end{abstract}

\section{Introduction}
%\todo{To rephrase: In~\cite{grivet2021speed}, the cryptosystem was tuned in order to have a very accurate argmax and, afterwards, the results were noised to achieve DP. In this present work, we aim at avoiding this pointless and costly (peut-être un peu rude)  return journey in accuracy by designing an homomorphic argmax whose imperfection leads to both a light computational cost and DP protection.}

As a protocol for training neural network without explicit sharing of the learning data, the Private Aggregation of Teacher Ensembles (PATE) approach has received much attention since its inception in the seminal work of Papernot et al \cite{papernot2016semi}. %In a nutshell, after an initial unsupervised training phase of a student model, the PATE protocol builds a labeled data set which is then used within a subsequent supervised training step which finalizes the student model. 
In a nutshell, the PATE protocol labels a subset of a public dataset and uses this partially labeled dataset to train a student model in a semi-supervised way.
The labelization is achieved by aggregating, usually by means of majority voting, the labels - considered as votes - provided by a set of teachers which are the owners of private data sets. Since the teachers' labels would leak information on their training data, the PATE protocol makes use of differential privacy (DP). To get a reasonable privacy-utility trade-off, the vote aggregation is performed on an independent server, the single elected vote seen by the student model being much easier to sanitize than the full histogram of the votes.
%requires the vote aggregation to be performed on an independent server resulting in the student model seeing only single consolidated votes. Nevertheless,  However, it is now known that this baseline protocol is not sufficient to guarantee the privacy of the teachers' training data from the student (as even the consolidated labels result in significant leakage on these data), the server (as knowledge of the individual votes also result in the same) as well as any entity to which the final model may eventually be disclosed.

%As a consequence, more recent works have investigated how to improve the privacy guarantees of the protocol. In order to do so, two main tools have been investigated. First, Differential Privacy (DP) noise is applied on the votes histogram before aggregation. This allows to better control the information leakage that can be exploited by both the student (which now sees only perturbated consolidated votes) and by the entities to which the final model is disclosed (DP guarantees being monotonous by postprocessing).
Still, in such a setting, the server has to be trusted since it sees the clear votes sent by the teachers. This is why SPEED \cite{grivet2021speed} builds upon the work from \cite{papernot2016semi} and uses fully homomorphic encryption (FHE) to blind the server by having it performing the aggregation directly over encrypted votes, therefore with neither knowledge of the individual votes nor of the consolidated one. In that work the authors associate a distributed Laplacian noise generation mechanism and carefully crafted homomorphic histogram and argmax computations. Still, FHE being computationally intensive, this comes at significant communication and computation costs on the server ($6.5$ minutes to compute the homomorphic argmax for $100$ queries).

In this paper, we revisit the association of DP and FHE in a radically different fashion. Indeed, rather than proceeding in two steps (noise addition and then homomorphic aggregation) we proceed by designing a new aggregation algorithm which has the desirable property of being much more efficient to evaluate over FHE but the less desirable property of being (stochastically) inaccurate. We then demonstrate that the inaccuracies of our algorithm translate into consistent DP guarantees, and therefore that explicit noise addition becomes unnecessary for DP. In doing so, and by means of a carefully crafted FHE implementation of the algorithm, we are able to achieve a reduction of $20\%$ in the computational burden of the aggregation server compared to the state of the art (\cite{IZ21}). To the best of our knowledge, it is the first work in which relaxing the accuracy of an homomorphic calculation is constructively usable as a degree of freedom to achieve better FHE performances. 

The paper is organized as follows. First of all, we explore the related work in Section~\ref{sec:related_work} and remind some preliminaries about HE and DP in Section~\ref{sec:preliminaries}. Then, we introduce and describe our argmax operator SHIELD in Section~\ref{sec:shield} and more specifically its FHE implementation in Section~\ref{sec:fhe_implementation}, before presenting SPEED application case in Section~\ref{sec:speed_application_case}. Section~\ref{sec:shield_analysis} develops an analysis of SHIELD from the points of view of DP and HE. Finally, our experimental results are presented in Section~\ref{sec:shield_experiments}.

\section{Related work}
\label{sec:related_work}
%DP, cryptographic primitives, secure aggregation.
%Private voting rules ?
%\todo{Function secret sharing : add categorical data with smaller vectors than one-hot encodings (distributed point function) \cite{boyle2015function}.}

%\subsection{Noisy cryptographic implementations ensuring differential privacy}
In \cite{wagh2021dp}, the authors survey recent works in which DP and cryptographic primitives take advantage of each other, either
\begin{itemize}
    \item cryptography for DP: cryptographic primitives allow to get the privacy-utility trade-off of a standard DP mechanism but without the need of a trusted server \cite{agarwal2018cpsgd, goryczka2015comprehensive, cheu2019distributed, erlingsson2019amplification}. This is an improvement compared to \emph{local DP} which, by making the data owners noise their data before outsourcing them, does not need a trusted server either but gives a poorer privacy-utility trade-off \cite{ullman2018tight, kasiviswanathan2011can}
    \item or DP for cryptography: design ``leaky'' cryptographic primitives that ensure DP and are more efficient than traditional primitives \cite{bater2018shrinkwrap, van2015vuvuzela, wagh2018differentially}
\end{itemize}
\cite{bater2018shrinkwrap, van2015vuvuzela, wagh2018differentially} are tailored to specific applications, respectively SQL queries, anonymous communication systems and oblivious RAM. Our work follows this line of DP for cryptography but in the context of election.
%They speak about the trade-off between privacy, utility and performance (performance is added due to the complexity of cryptographic primitives).

In \cite{zhu2020federated}, the authors propose an algorithm with a close goal, namely heavy-hitters (most frequent items) detection, which is inherently differentially private thanks to random sampling. Nevertheless, the goal of this inherent probabilistic behavior is not computational efficiency since the method is not articulated with cryptographic primitives. Moreover, this algorithm works on sequential data. Even if it does not restrict its generality since any data can be seen as sequential, the utility does depend on the sequential representation of the data, which may not be optimal if there is no semantic value to this representation. Finally, the algorithm is iterative and thus requires a lot of communication with the users.

As far as federated learning is concerned, the work from \cite{stevens2022efficient} is interesting because it leverages the error induced by encryption to derive DP guarantees. The aggregation protocol is based on the security of LWE problem and on the Multi-Party Computation protocol of Packed Shamir secret sharing scheme \cite{matthew1992}. Nevertheless, LWE is not used to directly encrypt the values of interest but rather to generate one-time pads of the same dimension of these values while only needing to communicate much smaller vectors to the server. These one-time pads allow a secure aggregation and DP guarantees are ensured by the error induced by LWE encryption.% We remark however that the communication and computation complexities are quite high.
%We remark however that the costs for communications are quite high with a communication complexity of $O(m+k+n)$ per client and for the server a communication complexity of $O(mk+n)$ with $m$, the size of the noised gradients, $k$ the number of clients and $n$ the number of shares (the size of the secret vector). Also, the computation complexity remains important with, on the client side, a complexity of $O(m n+k log(k))$ and on the server side of $O(m k+m n+k log(k))$.

\section{Preliminaries}
\label{sec:preliminaries}
\subsection{Homomorphic encryption}
Fully homomorphic encryption (FHE) schemes allow to perform arbitrary computations directly over encrypted data. That is, with a fully homomorphic encryption scheme $\mathsf{Enc}$, we can compute $\mathsf{Enc(m_1+m_2)}$ and $\mathsf{Enc(m_1\times m_2)}$ from encrypted messages $\mathsf{Enc(m_1)}$ and $\mathsf{Enc(m_2)}$. 

In this section we recall the general principles of the  \textbf{BFV} homomorphic cryptosystem \cite{fan2012somewhat} which will be used in a batched manner. Since we know in advance the function to be evaluated homomorphically, we can stick to the somewhat homomorphic version described below. % to reformulate
Let $R=\mathbb{Z}\left[x\right]/\Phi_m\left(x\right)$ denote the polynomial
ring modulo the $m$-cyclotomic polynomial  with $n=\varphi(m)$. The ciphertexts in the scheme are elements of polynomial ring $R_{q}$,
where $R_{q}$ is the set of polynomials in $R$ with coefficients
in $\mathbb{Z}_{q}$. The plaintexts are polynomials belonging to the ring $R_t=R/tR$. For $a \in R$, we denote by $\left[a\right]_q$ the element in R obtained by applying modulo q to all its coefficients.   
As such, \textbf{BFV} scheme is defined by the following probabilistic polynomial-time algorithms:

\textbf{BFV.ParamGen($\lambda$)} $\rightarrow (n, q, t, \chi_{key} , \chi_{err}, w)$.
It uses the security parameter $\lambda$ to fix several other parameters such as  $n$, the degree of the polynomials,  the ciphertext modulus $q$, the plaintext modulus $t$, the error distributions, etc.
	
\textbf{BFV.KeyGen$(n, q, t, \chi_{key} , \chi_{err}, w)$} $\rightarrow (pk,sk,evk)$. Taking as input the parameters generated in \textbf{BFV.ParamGen}, it calculates the private, public and evaluation key. Besides the public and the private keys, an evaluation key is generated to be used during computation on ciphertexts in order to reduce the noise.

\textbf{BFV.Enc$_{pk}(m)$}$\rightarrow c=(c_0,c_1)$. 
%For $m \in R_t$, sample $u\leftarrow \chi_{key}$, $e_1,e_2\leftarrow \chi_{err}$ and compute the ciphertext $c=([\Delta[m]_t+bu+e_1]_q, [au+e_2]_q)\in R_q^2$.
For $m \in R_t$, compute the ciphertext $c=(c_0,c_1) \in R_q^2$, using the public key $pk$. 

\textbf{BFV.Dec$_{sk}(c)$}$\rightarrow m$. It computes the plaintext $m$ from the ciphertext $c$, using private key $sk$.
	
%	\textbf{BFV.Eval}$_{pk,evk}(f,c_1,\ldots,c_n)$:$\rightarrow c$,  with  $c = $\textbf{BFV.Enc}$_{pk}(f(m_1,\ldots,m_n)) $, where $c_i=$\textbf{BFV.Enc}$_{pk}(m_i)$, and $f$ has $n$ inputs and has degree at most two. 
%	\\It allows the homomorphic evaluation of $f$, gate-by-gate over $c_i$  using the following functions: \textbf{BFV.Add$(c_1,c_2)$} and \textbf{BFV.Mul$_{evk}(c_1,c_2)$}.
\textbf{BFV.Add$(c_1,c_2)$}$ \rightarrow c_{add}$ with $c_{add}=\left( \left[c_{1,0}+c_{2,0}\right]_q,\left[c_{1,1}+c_{2,1}\right]_q\right)$.
		
\textbf{BFV.Mul$_{evk}(c_1,c_2)$} $\rightarrow c_{mul}=(c_0,c_1,c_2)$
with $c_0= \left[\left\lfloor{\frac{t}{q} \cdot c_{1,0}\cdot c_{2,0} }\right\rceil \right]_q$, $c_1=\left[\left\lfloor{\frac{t}{q} \cdot \left(c_{1,0}\cdot c_{2,1}+c_{1,1}\cdot c_{2,0} \right )}\right\rceil \right]_q$ and $c_2=\left[\left\lfloor{\frac{t}{q} \cdot c_{1,1}\cdot c_{2,1}}\right\rceil \right]_q$.
%It computes  \\$c_{mul}=\left(\left[\left\lfloor{\frac{t}{q} \cdot c_{1,0}\cdot c_{2,0} }\right\rceil \right]_q, \left[\left\lfloor{\frac{t}{q} \cdot \left(c_{1,0}\cdot c_{2,1}+c_{1,1}\cdot c_{2,0} \right )}\right\rceil \right]_q,\left[\left\lfloor{\frac{t}{q} \cdot c_{1,1}\cdot c_{2,1}}\right\rceil \right]_q\right)$  

In order to reduce the number of elements in the ciphertexts obtained after a multiplication, a relinearization method is proposed:
\textbf{BFV.Rel$(c_0,c_1,c_2)$} $\rightarrow ct'=(c'_0,c'_1)$ such that $\left[c_0+c_1 * sk + c_2 * s^2 \right]_q=\left[c'_0+c'_1*sk + r \right]_q$ with the norm $||r||$ small. 

For further details on the precise two relinearization methods and the full description of the scheme, we refer the reader to the original paper \cite{fan2012somewhat}.
Let us also note that to this original scheme, one can apply batching (also known as packing), an optimization method for FHE allowing to put several clear messages into a single ciphertext and execute parallel operations on them into a SIMD (Single Instructions Multiple Data) manner. The technique of ciphertext-packing is based on polynomial CRT (Chinese Reminder Theorem) and was originally described in \cite{smart2011}, \cite{brakerski2014}. 
%add refs
%to be completed if more details are needed
% add TFHE ?
% add details about batching since we will use it ?

\subsection{Differential privacy}

Differential privacy~\cite{dwork2006our} is a gold standard concept in privacy-preserving data analysis. It provides a guarantee that under a reasonable privacy budget $(\epsilon,\delta)$, two adjacent databases produce statistically indistinguishable results. In this work, we consider that two databases $d$ and  $d'$ are adjacent if they differ by at most one example.
\begin{definition}
   A randomized mechanism $\RandMech$ with output range $\mathcal{R}$ satisfies \emph{$(\epsilon,\delta)$-DP} if for any two adjacent databases $d$ and $d'$ and for any subset of outputs $S\subset \mathcal{R}$ one has
   \begin{equation*}
       \mathbb{P}\left[ \RandMech(d) \in S\right] \leq e^\epsilon \mathbb{P}\left[\RandMech(d') \in S\right] +\delta.
   \end{equation*}
\end{definition}

\begin{definition}
Let $\RandMech$ be a randomized mechanism with output range $\mathcal{R}$ and $d$, $d'$ a pair of adjacent databases. Let $\aux$ denote an auxiliary input. For any $o\in \mathcal{R}$, the \emph{privacy loss} at $o$ is defined as 
\begin{equation*}
    c(o;\RandMech,\aux,d,d'):= \log \left( \frac{\mathbb{P}[\RandMech(\aux,d) =o]}{\mathbb{P}[\RandMech(\aux,d') =o]} \right).
\end{equation*}

We define the \emph{privacy loss random variable} $C(\RandMech,\aux,d,d')$ as
\begin{equation*}
    C(\RandMech,\aux,d,d') := c(\RandMech(d);\RandMech,\aux,d,d')
\end{equation*}
\emph{i.e.} the random variable defined by evaluating the privacy loss at an outcome sampled from $\RandMech(d)$.
\end{definition}

We determine the privacy loss of our protocol via a two-fold approach. First of all, we determine the privacy loss per query and then we compose the privacy losses of each query to get the overall loss. The classical composition theorem (see e.g.~\cite{dwork2014algorithmic}) states that the guarantees $\epsilon$ of sequential queries add up. Nevertheless, training a deep neural network, even with a collaborative framework as presented in this paper, requires a large amount of calls to the databases, precluding the use of this classical composition. Therefore, to obtain reasonable DP guarantees, we need to keep track of the privacy loss with a more refined tool, namely the moments accountant~\cite{abadi2016deep} whose definition we recall here.

\begin{definition}
\label{def:moments_accountant}
With the same notations as above, the \emph{moments accountant} is defined for any $l\in \mathbb{N_+^*}$ as
\begin{equation*}
    \alpha_{\RandMech}(l):=\max_{\aux,d,d'}\alpha_{\RandMech}(l;\aux,d,d') 
\end{equation*}
where the maximum is taken over any auxiliary input $\aux$ and any pair of adjacent databases $(d,d')$ and $\alpha_{\RandMech}(l;\aux,d,d'):=\log \left(\mathbb{E}\left[\exp(l C(\RandMech,\aux,d,d'))\right]\right)$ is the moment generating function of the privacy loss random variable.
\end{definition}

The two following properties of the moments accountant allow to compose the privacy costs along the queries and then go back to the standard DP guarantee. Composing in the moments accountant framework often yields far better DP guarantees in the end.

\begin{proposition}[\cite{abadi2016deep}]
     \label{prop:moments_composition}
      Let $p\in \mathbb{N}^*$. Let us consider a mechanism $\RandMech$ defined on a set $\mathcal{D}$  that consists of a sequence of adaptive mechanisms $\RandMech_1, \dots, \RandMech_p$ where, for any $i\in \{1, \dots, p\}$, $\RandMech_i\colon \prod_{j=1}^{i-1}\mathcal{R}_j\times \mathcal{D} \mapsto \mathcal{R}_i$. Then, for any $l\in \mathbb{N^*}$,
    \begin{align*}
        \alpha_{\RandMech}(l) \leq \sum_{i=1}^p \alpha_{\RandMech_i}(l).
    \end{align*}
\end{proposition}

\begin{proposition}[\cite{abadi2016deep}]
    \label{prop:tail_bound}
    For any $\epsilon \in \mathbb{R}_+^*$, the mechanism $\RandMech$ is $(\epsilon, \delta)$-differentially private for $\delta = \min_{l\in \mathbb{N}^*} \exp(\alpha_{\RandMech}(l) - l\epsilon)$.
\end{proposition}

Finally, an important property of DP, widely used in DP analysis, is that it is immune to post-processing.

\begin{proposition}[\cite{dwork2014algorithmic}]
	\label{prop:post_processing}
	Let $\RandMech$ be a probabilistic mechanism, with output range $\mathcal{R}$, that is $(\epsilon, \delta)$-differentially private, with $(\epsilon, \delta)\in \left(\mathbb{R_+}\right)^2$. Let $\phi \colon \mathcal{R} \to \mathcal{R'}$ be an arbitrary probabilistic mapping. Then $\phi \circ \RandMech$ is $(\epsilon, \delta)$-differentially private.
\end{proposition}

\section{SHIELD: Secure and Homomorphic Imperfect Election via Lightweight Design}
%\subsection{The tractable Confidential Light Argmax With Sampling (tractable CLAWS)}
\label{sec:shield}
In this paper, for any $m\in \mathbb{N}$, $[m]$ will denote the set $\{1,..., m\}$ (which is, by convention, the empty set if $m=0$).

Let $K$ be the number of classes of the classification problem.
Let $n$ be the number of voters or teachers and, given a sample and $k\in [K]$, let $n_k$ be the number of teachers who voted for class $k$.

\subsection{Principle of SHIELD}
\label{sec:principle_shield}
We propose a novel operator that can be viewed as an aggregation operator for categorical data, as well as a voting rule, or even a probabilistic argmax. This operator, called SHIELD (Secure and Homomorphic Imperfect Election via Lightweight Design) aims at computing the aggregation of categorical data - or equivalently the winner of an election - on a server while ensuring the privacy of the inputs from both the server and the end-users that may try to retrieve sensitive information from the output. Let us now formally introduce SHIELD.

First of all, SHIELD is meant to be computed in the homomorphic domain. Here are some notations we will use to describe its homomorphic behavior. $\enc$ and $\dec$ respectively denote the encryption and decryption functions of some homomorphic encryption system defined on $\mathbb{Z}_2$. $\oplus$ and $\otimes$ respectively represent the homomorphic addition and multiplication. When these operators are applied on vectors, they denote the element-wise corresponding operations. Note that the negation of $x\in \mathbb{Z}_2$ is homomorphically performed via $\enc(1) \oplus \enc(x)$ and the homomorphic \emph{or} operator, denoted $\olor$, between $x\in \mathbb{Z}_2$ and $y\in \mathbb{Z}_2$ is performed via $[\enc(x) \oplus \enc(y)] \oplus [\enc(x) \otimes \enc(y)]$ and will be written $\enc(x) \olor \enc(y)$ in the following.

\begin{definition}
Let $K\in \mathbb{N}^*$. A vector $z\in (\mathbb{Z}_2)^K$ is said to be a \emph{one-hot encoding} vector if there exists $k_0\in [K]$ such that $z_{k_0} = 1$ and, for all $k\in [K] \setminus \{k_0\}$, $z_{k} = 0$. In this case, we say that $z$ \emph{codes} for the class $k_0$ or that $z$ is the one-hot encoding of the class $k_0$.
\end{definition}

Let $(p,a)\in (\mathbb{N}^*)^2$.
Let $\shield{p}{a}$ denote SHIELD operator with parameters $p$ and $a$, that we define in the following.

Let $(n, K)\in (\mathbb{N}^*)^2$ that we consider fixed in the remainder of this section.
Let $Z = (\enc(z^{(i)}))_{i\in [n]}$ be a list of $n$ encrypted one-hot encoding $K$-dimensional vectors, some of these vectors being possibly equal (it is necessarily the case for some vectors when $K < n$). Then $\shield{p}{a}(Z)$ is an encryption of one of the $z^{(i)}$, and with high probability (see Section~\ref{sec:shield_experiments} for quantitative results) $\shield{p}{a}(Z)$ is an encryption of the most frequent of the one-hot encoding vectors of $Z$. $\shield{p}{a}$ is formally defined in Algorithm~\ref{algo:shield} where, for the sake of clarity, we do not explicitly write the encryption function (e.g. $res$ = $z^{(i_0)}$ instead of $res$ = $\enc(z^{(i_0)})$). $\shield{p}{a}$ draws $p$ vectors of $Z$ \emph{with replacement} in a uniformly random manner and multiply them. The resulting vector $\pi$ is an encryption of the one-hot encoding of the class $k_0$, $k_0\in [K]$, if all the $p$ drawn encrypted vectors code for the same class $k_0$. Otherwise, $\pi$ is the null vector of $(\mathbb{Z}_2)^K$. If a non-null vector has already been found, the current $\pi$ is ignored (since the bit $found\_not\_null$ has been set to $1$). Of course, since the algorithm is computed in the encrypted domain, it has to run until the end of the \textit{for} loop but everything works as if the algorithm repeated this operation until it gets a non-null vector and then ignored the remaining product vectors. This first non-null vector is the output of $\shield{p}{a}$. If no non-null vector was produced after $a$ iterations, a null vector is output and we say that $\shield{p}{a}$ \emph{failed}.

\begin{algorithm}
	\SetAlgoLined
	\SetKwInOut{KwIn}{Input}
	\SetKwInOut{KwOut}{Output}
	\KwIn{number of vectors $n$, number of classes $K$, list of encrypted votes $Z$, number of multiplications $p$, number of terms $a$}
	\KwOut{$res$ = $z^{(i_0)}$ where $i_0\in [n]$}
	$res \leftarrow (0, \dots, 0)\in (\mathbb{Z}_2)^K$\;
	$found\_not\_null \leftarrow 0$\;
	\For{$j$ in $[a]$}{
	 $\pi \leftarrow (1, \dots, 1)\in (\mathbb{Z}_2)^K$\;
	 \For{$l$ in $[p]$}{
	  Draw a vector $z$ of $Z$ uniformly at random\;
	  $\pi \leftarrow \pi \otimes z$\;
	 }
	 $res \leftarrow res \oplus (1 \oplus found\_not\_null) \otimes \pi$\;
	 $is\_not\_null \leftarrow \bigoplus_{k=1}^K \pi_k$\;
	 $found\_not\_null \leftarrow found\_not\_null \olor is\_not\_null$\;
	}
	\caption{SHIELD}
    \label{algo:shield}
\end{algorithm}

$a$ being fixed, the choice of $p$ must consider the trade-off between, on one hand, the accuracy of the operator, e.g. the probability of getting the truly most frequent vector (see the considered accuracy metrics in Section~\ref{sec:accuracy}), and, on the other hand, the probability of avoiding a failure and the computational complexity. Indeed, when $p$ increases, the probability of getting a null vector (and then failing) increases, as well as the computational complexity, but the probability of getting the most frequent vector, knowing that the algorithm did not fail, increases too.

\subsection{Multi-degree SHIELD}
We can imagine a parameter $p$ that decreases as the iterations run, as if it adapted to the vote distribution. Indeed, on one hand, a high $p$ for the first iterations ensures (with high probability) that we get the truly most frequent vector if getting a non-null vector is easy (i.e. probable), which happens if a vast majority of the vectors code for the same class (e.g. a vast majority of voters agree on one candidate). On the other hand, if the first iterations failed, which suggests that getting a non-null vector is not so probable, the number $p$ of multiplications decreases in order to make the production of a non-null vector easier. In this framework, our SHIELD operator can be represented by a polynomial $\sum_{p=1}^D a_p X^p$ with positive integer coefficients, where $\sum_{p=1}^D a_p = a$ and some $a_p$'s may be null. We call $\sum_{p=1}^D a_p X^p \in \mathbb{N}[X]$ the \emph{polynomial parameterization} of SHIELD. There is indeed a bijection between the set of operators and $\mathbb{N}[X]$ since the order of the terms of different degrees is constrained to be the one of decreasing degrees. Nevertheless, the analogy seems to stop here since the algebraic structure of $\mathbb{N}[X]$ does not apply to the set of operators (think about a factorization like $X^2\sum_{p=0}^D a_p X^{p-2}$, that would draw for once two vectors and use them for all the $a$ terms, whereas we here want to independently draw the vectors for each term).

%To choose which polynomial to use, one could draw inspiration from the exponential mechanism~\cite{mcsherry2007mechanism} which must give good utility in expectation together with well-known DP guarantees. The development in series of the exponential function may indeed guide the choice of the polynomial coefficients.

Note that we can easily ensure that multi-degree SHIELD does not fail by imposing $a_1 = 1$. Indeed, when we draw only one one-hot encoding vector, without multiplying it with others, we cannot get a null vector. Moreover, $a_1 > 1$ is useless since the first draw of a single vector will succeed.

It is easily seen that multi-degree SHIELD is a generalization of SHIELD and, as such, in the remainder of this article, multi-degree SHIELD will simply be referred to as SHIELD.

\subsection{Offset parameter}
\label{sec:offset}
The SHIELD operator as defined above cannot always provide finite DP guarantees. Let us consider two adjacent databases $d$ and $d'$ such that, in $d$, a class $c$ was chosen by no voter and, in $d'$, $c$ was chosen by one voter. Then, with input $d$, SHIELD will never output $c$ because it cannot pick a one-hot encoding for $c$, the probability of outputting $c$ is then null. On the contrary, with input $d'$, there is a non-null probability (even if it is small) of outputting $c$. Hence, the ratio of probabilities of outputting $c$ is not bounded and we get an infinite privacy cost.

To avoid this problem, we force all the classes to have at least one vote by creating a dummy one-hot encoding for each class. More generally, $\offset$ dummy one-hot encodings can be created for each class, where $\offset$ is another parameter of SHIELD, called the \emph{offset}.

Algorithm~\ref{algo:multi_degree_shield} gives the pseudocode of the multi-degree version of SHIELD with the offset parameter.

\begin{algorithm}
	\SetAlgoLined
	\SetKwInOut{KwIn}{Input}
	\SetKwInOut{KwOut}{Output}
	\KwIn{number of vectors $n$, number of classes $K$, list of encrypted votes $Z$, polynomial $(a_p)_{p\in [D]}$, offset $\offset$}
	\KwOut{$res$ = $z^{(i_0)}$ where $i_0\in [n]$}
    $Z \leftarrow Z$ augmented by $\offset$ encrypted one-hot encodings for each class\;
	$res \leftarrow (0, \dots, 0)\in (\mathbb{Z}_2)^K$\;
	$found\_not\_null \leftarrow 0$\;
	\For{$p$ in $[D]$}{
		\For{$j$ in $[a_p]$}{
		    $\pi \leftarrow (1, \dots, 1)\in (\mathbb{Z}_2)^K$\;
		    \For{$l$ in $[p]$}{
    			Draw a vector $z$ of $Z$ uniformly at random\; \label{algline:teacher_selection}
    			$\pi \leftarrow \pi \otimes z$\; \label{algline:mult_of_pis}
			}
		    $res \leftarrow res \oplus (1 \oplus found\_not\_null) \otimes \pi$\; \label{algline:res_update}
    		$is\_not\_null \leftarrow \bigoplus_{k=1}^K \pi_k$\; \label{algline:sum_of_pis}
    		$found\_not\_null \leftarrow found\_not\_null \olor is\_not\_null$\; \label{algline:res_bit_update}
		}
	}
	\caption{Multi-degree SHIELD}
	\label{algo:multi_degree_shield}
\end{algorithm}

In our experiments, we fixed $\offset$ to $1$, letting the optimization of this parameter for further work. It is nevertheless intuitive that the greater $\offset$, the worse the accuracy because, when $\offset$ is large, the distribution of the votes is flattened and the probability of outputting the true argmax is lower.

\subsection{Exponential argmax operator}
\label{sec:exponential_argmax}
As an inherently stochastic mechanism that does not resort to noise addition but rather outputs a value with a probability that is an increasing function of its utility (if we deem that the vote frequency of a class constitutes its utility), SHIELD can be compared to the exponential mechanism (introduced in \cite{mcsherry2007mechanism}) which samples its output following the softmax distribution of the utility. However, the sampling in the encrypted domain constrains the shape of the probability distribution and introduces a dependency of the practically implementable distributions with the computational efficiency of the operator.

Note that softmax has been approximately implemented in FHE through polynomial approximation \cite{lee2022privacy} but this requires a quite high multiplicative depth (with a polynomial of degree 12 for approximating the exponential function and even more for approximating the inverse function) and results in a significant computational overhead. Moreover, using such an implementation would still require additional homomorphic operations like comparisons to actually sample the output according to this distribution.

Rather, a method of sampling that follows the exponential distribution by construction, in the spirit of SHIELD as presented in this paper, would be more seducing. Sampling each vote independently with a fixed probability would actually yield an output distribution that exponentially depends on the vote frequencies but it seems that the probability of failing by not outputting any class would be quite high for practical parameters. We let further work on this question as a perspective.

\section{FHE implementation of SHIELD}
\label{sec:fhe_implementation}

    Algorithm \ref{algo:multi_degree_shield} is a generic version of SHIELD that actually needs to be adapted for an implementation using an HE cryptosystem. And first, there are two kinds of possible encodings depending on the encryption scheme that is used:

    \begin{itemize}
        \item Single Instruction, Multiple Data (SIMD). Using the BFV cryptosystem, a number of values are encoded simultaneously in a polynomial which is then encrypted. A single operation on a ciphertext leads to the same operation applied to all values encoded inside the ciphertext.
        \item Single Instruction, Single Data (SISD). One way of using the TFHE cryptosystem is to use a single ciphertext to encrypt a single value. This is less efficient than using SIMD but unlocks a set of complex operations on that ciphertext that are impossible to implement otherwise.
    \end{itemize}

    We implement SHIELD with two separate methods: one uses the BFV cryptosystem with SIMD operations; the other uses the TFHE cryptosystem with SISD operations.

    \subsection{Implementing SIMD-SHIELD}
    \label{subsec:SIMD-SHIELD-implementation}

    Although using BFV allows us to speed up SHIELD considerably by batching different samples together in the same ciphertext, some constraints require adapting parts of Algorithm \ref{algo:multi_degree_shield} for them to work. \\

    \textbf{a. Multiplicative depth.} As it is the case for other similar HE schemes, we need to set the parameters of BFV according to the multiplicative depth of the computation. The higher the multiplicative depth, the bigger the parameters, and the less efficient the overall computation. For this reason, some parts of the algorithm, like Line \ref{algline:mult_of_pis}, need to be changed. We can store all of the values for $\enc(z)$ over the loop and multiply them in a classic tree-based approach (instead of multiplying them sequentially) which reduces the multiplicative depth of the computation from $p$ to $\log_2(p)$. 

    The same change is applied everywhere it is needed, that is to say at Lines \ref{algline:res_update} and \ref{algline:res_bit_update} of Algorithm \ref{algo:multi_degree_shield}. \\

    \textbf{b. Selecting the teacher.} Selecting the voter, also called teacher because of SPEED application case (see Section~\ref{sec:speed_application_case}), at lines \ref{algline:teacher_selection} and \ref{algline:mult_of_pis} of Algorithm \ref{algo:multi_degree_shield} is easy enough when the SHIELD algorithm is called for a single sample at once. However, in order to speed up the algorithm and make use of the SIMD property of the BFV cryptosystem fully, we actually run the SHIELD algorithm for a number of samples at a time.

    For instance, if $\pi^{(i)}$ is the $\pi$ vector of $K$ values for sample $i$, then the actual vector encoded in the ciphertext for the packed algorithm would be 

    \begin{equation}
         \label{eq:packed-pis}
         \pi = \left(\pi_1^{(1)}, \dots, \pi_K^{(1)}, \pi_1^{(2)}, \dots, \pi_K^{(2)}, \dots \right)
    \end{equation}

    This allows us to use the full size of the polynomials we encrypt. These polynomials have degrees in the order of $\approx 2^{15}$ while $K$ is usually in the order of $\approx 10$.

    Therefore the teacher selection step has to be modified. The new encoding of teachers $t$'s vote for sample $i$ is:

    \begin{equation*}
        \left(0, \dots, 0 | 0, \dots, 0 | \dots | z^{(i)}_t | \dots | 0, \dots, 0 | \right) \in \B^{N \times K}
    \end{equation*}

    which is a vector with $N$ slots of $K$ binary values where $z^{(i)}_t$ is teacher $t$'s original one-hot encoded vote for sample $i$. It is located at the $i^{\text{th}}$ slot of the encoding. From now on we'll call $z^{(i)}_t$ this new encoding of the teacher's votes. Algorithm \ref{algo:packed-teacher-selection} presents the process for teacher selection and creation of the $\pi$ vector using this new encoding.
    
    \begin{algorithm}
    	\label{algo:packed-teacher-selection}
    	\SetAlgoLined
        \For{$j$ in $[a_p]$}
        {            \label{algline:poly_round_1}
            \For{$l$ in $[p]$}
            {   \label{algline:poly_round_2}    
                \For{$t$ in $[n]$}
                {
                    $z_t \leftarrow (0, \dots, 0)$\;
                }
                \For{$i$ in $[N]$}
                {
                    Draw a vector $z^{(i)}_t$ of $Z$ uniformly at random\; 
                    $z_t \leftarrow z_t \oplus z^{(i)}_t$ \; 
                    update $m_t$ \;     \label{algline:mask_update}
                }

                $\pi \leftarrow (1, \dots, 1)$\;
                \For{$t$ in $[n]$}
                {
                    $z_t \leftarrow z_t \oplus m_t$ \;     \label{algline:mask_addition}
                    $\pi \leftarrow \pi \otimes z_t$ \;
                }
            }
        }
    	\caption{Teacher selection. With $n$ the total number of teachers, this algorithm describes the actual steps for selecting the teachers that get to vote in the SIMD encoding paradigm.}
    \end{algorithm}

    At step \ref{algline:mask_update} a mask $m_t$ is updated but no detail is given for clarity. For every teacher $t$, the mask $m_t$ is a plaintext vector that contains $0$s in the place of samples for which the teacher votes and $1$s in the place of samples for which the teacher does not vote.
    As an example, for $K=2$ and $N=4$, if teacher $t$ votes for samples $1$ and $3$, then

    \begin{equation*}
        m_t = \left(0, 0 | 1, 1 | 0, 0 | 1, 1 \right)
    \end{equation*}

    This mask is then added to $z_t$ before the multiplication to the $\pi$ vector so that all the samples that are not voted on do not impact the result: their slots are filled by ones. If the mask is not used, then all non-selected slots will be filled with 0s and therefore would set everything to 0 after the multiplication.
    
    For this multiplication, as mentioned before, we opt to store all of the $z_t$ vectors and create a multiplication tree to reduce the multiplicative depth. \\

    \textbf{c. Rotations.} One other constraint that schemes such as BFV suffer from, is that it is very hard and costly to extract certain values from the ciphertext to apply an operation \emph{only} to them. Such is the case when trying to implement Line \ref{algline:sum_of_pis} in Algorithm \ref{algo:multi_degree_shield}. The individual $\pi_k$ values cannot be extracted and summed together in a straight-forward manner. One thing we can do however, at a relatively low cost (both in terms of performance and noise inside the ciphertext), is to rotate the vector encoded in the ciphertext. This leads to an implementation of Line \ref{algline:sum_of_pis} that we present using the example $\pi^{(i)} = \left(0,0,0,1,0,0,0,0,0,0\right)$.

    \begin{align*}
            & \left( 0,0,0,1,0,0,0,0,0,0 \right) & \\
        +   & \left( 0,0,0,0,0,?,?,?,?,? \right) & \leftarrow \text{rotate by 5} \\
        =   & \left( 0,0,0,1,0,?,?,?,?,? \right) & \\
        +   & \left( 0,1,0,?,?,?,?,?,?,? \right) & \leftarrow \text{rotate by 2} \\
        =   & \dots &
    \end{align*}

    One can see how, using $\log_2(K)$ rotations and sums, we can obtain $\sum_{j} \pi^{(i)}_j$ in the \emph{first} coordinate of the $\pi^{(i)}$ vector. The question marks $?$ represent values that are rotated over from the next slot, (recall the complete form of $\pi$ in equation \ref{eq:packed-pis}).
    
    Therefore, we cannot control the values in the rest of the coordinates. And this is not enough. For Line \ref{algline:res_bit_update} to work, we need to have a vector where \emph{all} coordinates $\pi^{(i)}_j$ are filled with $\sum_{j} \pi^{(i)}_j$, not just the first one. To obtain this, we have to multiply by a plaintext with values $(1,0,0,\dots)$ to select only for the first coordinate of $\pi$ and then re-populate the rest of the coordinates using rotations and sums exactly in the opposite way as used for the computation of the sum of the $\pi^{(i)}_j$ values. \\

    \textbf{d. Packing the polynomial rounds together.} Up until now, for clarity, we presented a version of our algorithm that packed all or some of the $N$ samples together in a single ciphertext. In practice, to speed up the computation further, we also pack the polynomial rounds together. What we mean by "polynomial rounds" is the two \emph{for} loops at Lines \ref{algline:poly_round_1} and \ref{algline:poly_round_2} in Algorithm \ref{algo:packed-teacher-selection}. We can remove these \emph{for} loops and compute them in parallel in a single ciphertext.

\section{An application case: SPEED}
\label{sec:speed_application_case}
\subsection{SPEED workflow}
\begin{figure}
    \centering	\includegraphics[width=\textwidth]{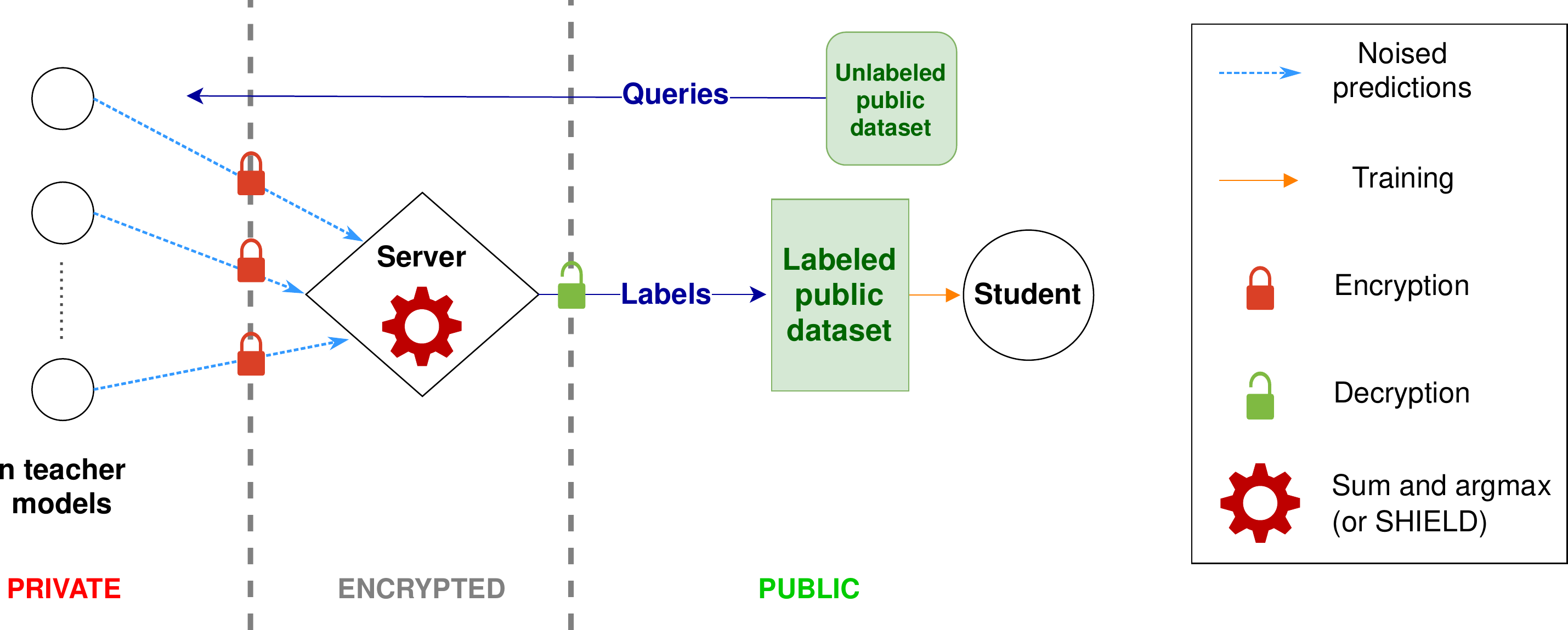}
    \caption{SPEED learning protocol}
    \label{fig:speed}
\end{figure}

Our SHIELD operator is actually tailored to a learning protocol called SPEED, from~\cite{grivet2021speed}, itself inspired from PATE~\cite{papernot2016semi}. SPEED method is illustrated by Figure~\ref{fig:speed}, inspired from~\cite{grivet2021speed}. Assuming the existence of a public unlabeled database $\publicD$ (we will keep this notation throughout the paper), SPEED enables several data-owners, called \emph{teachers}, to collaboratively train a classification model without outsourcing their data that are considered private. The idea is to label $\publicD$ and use it to train the final classification model, called the $\emph{student}$ model or simply the student. To do so, each teacher is asked to train a model beforehand for the same task as the student's target task with its own data only and, for each sample of $\publicD$ to label, every teacher infers a label through its model and sends this label to an aggregation server. The server then counts the number of labels received for each class, also seen as votes, and outputs the dominant class which is sent to the student for training.

As it was described, the protocol does not protect the data from the server or the end-users. Before explaining how data privacy is ensured, let us present the threat model.

\subsection{Threat model}
All the actors of the protocol, namely the teachers, the server and the student are considered honest-but-curious. This means that they execute their task correctly but may use the data they have access to to retrieve sensitive information about the teachers' data.
The end-users are also considered curious, the honest part not being relevant for end-users that are not involved in the training. Note that, in many real-life cases, the teachers may be end-users of the student model.

A limitation to the threat model is that the server is not considered to have access to the trained student model since our DP analysis assumes that the adversary only sees the output class, which is not exactly the case of the server (see \ref{sec:DP} for more details).

\subsection{Data protection}
To prevent the student and \textit{a fortiori} the end-users (by post-processing) from discovering sensitive information by attacks such as e.g. model inversion or membership inference, we apply DP. The teachers noise their votes before sending them to the server.

One could argue that the noise added by the teachers would also blur the sensitive information to the server. Nevertheless, the added noise is precisely scaled so that it protects the output of the aggregation, i.e. the dominant class, without harming too much the student accuracy. If the individual votes sent to the server were to be protected by DP \emph{before} aggregation, thus achieving what is called \emph{local DP} (\cite{kasiviswanathan2011can, duchi2013local, kairouz2016extremal}), this would require much more noise, too much noise to ensure a reasonable accuracy for the student model. As a consequence, the votes need to be protected from the server another way. This is where homomorphic encryption makes its entrance. After noising their votes, the teachers encrypt them. The server then receives the encrypted votes and perform their aggregation (sum and argmax) in the homomorphic layer. Finally, the output of the aggregation is sent to the student that owns the decryption key and is therefore able to decrypt it.

By the honest-but-curious hypothesis, we then assume that:
\begin{itemize}
    \item the teachers send their votes correctly noised and encrypted to the server
    \item the server performs the aggregation in the homomorphic domain as it is asked to
    \item the student decrypts the data and get trained; importantly, it does not share the decryption key with the server.
\end{itemize}

A real-life scenario could involve hospitals that own patients' medical data and aim at training a global model that would help the early diagnosis of a specific disease. In this case, the end-users would be the hospitals themselves.
%\todo{To  develop maybe}

\subsection{Faster SPEED with SHIELD}
Our SHIELD operator can be used to replace the sum and argmax computations on the server side in SPEED (represented by the gear wheel in Figure~\ref{fig:speed}). After receiving all the votes from the teachers, the server randomly picks some vectors with replacement as described in Section~\ref{sec:principle_shield}. Note that, being honest-but-curious, the server is trusted to compute SHIELD without mistake. Interestingly, the rest of SPEED protocol remains unchanged, except the sending of dummy one-hot encodings by some teachers, according to the offset parameter (see Section~\ref{sec:offset}).

%\subsection{Other application case: the vote}
%\todo{Say that SHIELD can be used more generally to compute the winner of an election without leaking any information about the voters to anyone. Look for references about private voting.}

\section{Analysis of SHIELD}
\label{sec:shield_analysis}

\subsection{\textit{A priori} accuracy metrics}
\label{sec:accuracy}
The ultimate accuracy that we want to maximize in SPEED application case is obviously the testing accuracy of the student model. Nevertheless, it could be interesting to measure the accuracy of the argmax operator itself, independently of the student training. Also, even if this depends on the teachers' votes and thus on the used dataset, this enables us to evaluate polynomial parameterizations without performing the student training, which is much faster and allows to test much more parameterizations. We call such an accuracy an \emph{a priori} accuracy.

The most straightforward way to define the argmax accuracy is probably to consider the probability of getting the exact argmax. % binary approach
Nevertheless, this approach treats any mistake the same way. It could be argued that outputting, say, the class that received the second greatest number of votes is better than outputting the least preferred class. Taking such a concern into account in our metric would also give a better hint about the student accuracy since, while the most preferred class (i.e. the exact argmax) is not always the ground truth class, a class with a lot of votes is more likely to be the ground truth class.

We could then make the assumption that the frequency of votes for a class is proportional to the probability of this class being the ground truth class of the sample (which is not necessarily the most preferred class). This would correspond to an assumption of well-calibrated vote distributions. In this context, another accuracy metric would be the probability of outputting the ground truth class of the sample. We call this metric the \emph{ground truth accuracy}, since it does not focus on outputting the exact argmax but rather the ground truth class. If $p_k$ denotes the probability of SHIELD outputting class $k$, for $k\in [K]$, the ground truth accuracy, written $\gta$, is:
\begin{align*}
    \gta = \sum_{k=1}^K \frac{n_k}{n}p_k.
\end{align*}
%\todo{To explain ?}

Of course, both metrics must be averaged on all the samples sent to the teachers.

\subsection{Differential privacy analysis}
\label{sec:DP}
Since the student model training requires many requests to the teachers and, indirectly, to their private datasets, we use, as in~\cite{grivet2021speed}, the moments accountant technique~\cite{abadi2016deep} to get a better privacy cost over composition.

We here consider that two databases $d$ and $d'$ are adjacent if they are the concatenations of the datasets from the same number of teachers and only one teacher differs from one database to the other. This implies that either all the $n'_k$, counts for database $d'$, for $k\in [M]$, are equal to the $n_k$, counts for database $d$, in which case the corresponding moments accountant is null, or the $n'_k$ differ from the $n_k$ only for two values of $k$, say $k_1$ and $k_2$, such that $n'_{k_1} = n_{k_1} - 1$ and $n'_{k_2} = n_{k_2} + 1$ (i.e. the differing teacher votes for $k_1$ in $d$ and $k_2$ in $d'$).

The stochastic behavior of our operator uncommonly does not come from an additional random noise, since the operator is inherently probabilistic. This is this very property of our operator that we leverage to ensure DP.
Computing the privacy cost of the training, as well as the \textit{a priori} accuracy, thus requires knowing the probabilities of outputting each class.

\subsection{Computing the probability distribution of the output}
We compute the probability distribution of the output of the algorithm SHIELD with a given polynomial parameterization in a recursive manner.

For a sample $x$ of $\publicD$, let $\randMech{P}{x}$ be the mechanism that takes the whole database (concatenation of the teachers' datasets) as input and outputs the class sent to the student i.e. the output of SHIELD, with the polynomial parameterization $P\in \mathbb{N}[X]$.

Let $d$ be the database composed of the teachers' data. Let $k$ be a class of the problem.

If $P = X$, $\mathbb{P}[\randMech{P}{x}(d) = k] = \frac{n_k}{n}$.

If $P = X^p + Q(X)$, where $Q\in \mathbb{N[X]}$ and $p\in \mathbb{N^*}$ is greater or equal than the degree of $Q$,
\begin{align*}
    &\mathbb{P}[\randMech{P}{x}(d) = k] = \left(\frac{n_k}{n}\right)^p + \left(1 - \sum_{j=1}^K\left(\frac{n_j}{n}\right)^p\right)\mathbb{P}[\randMech{Q}{x}(d)=k].
\end{align*}

Using these expressions, we simply compute the moments accountant for each query by taking the maximum over all pairs $(d, d')$ such that $d$ is the database constituted by the concatenation of the teachers' database and $d'$ is a database adjacent to $d$. We then derive the overall privacy cost using Propositions~\ref{prop:moments_composition} and \ref{prop:tail_bound}.

Note that the obtained DP guarantees are \emph{data-dependent} since we explored only the pairs of adjacent databases such that one of them is the actual database given by our application. The very values $\epsilon$ and $\delta$ of these guarantees then reveal some information about the training data. In a real-life scenario, these values should be sanitized before being published, as in~\cite{papernot2018scalable} for instance, but this is beyond the scope of this work.

\subsection{The differential privacy analysis does not apply to the server}
\label{sec:dp_not_for_server}
When we compute the probabilities of outputting a class, we do not suppose anything about whose votes are drawn i.e. we do not condition the probabilities on some particular drawing event. This amounts to assume that the adversary only sees the output class, and does not know, in particular, which teachers were selected in the sampling. This assumption cannot apply to the server since it draws the one-hot encodings itself and knows which teacher they come from, for having receiving the encodings one by one from the teachers.

To give an insight of why this subtlety is problematic, let us propose some concrete situations where the DP guarantees are obviously not protecting the vote of the server's victim, i.e. the teacher whose vote the server wants to know.
\begin{itemize}
    \item With the polynomial parameterization $X^k + X$, $k\in \mathbb{N}*\setminus \{1\}$, if the server draws $k-1$ teachers and its victim for the term $X^k$ and then its victim for the term $X$, then the server will know that the class sent to the student is its victim's vote.
    \item Supposing that the server knows the votes of all the teachers except its victim's (classical assumption in DP), it will be able to recover its victim's votes in many cases. For instance, with the polynomial parameterization $X^k + X$, $k\in \mathbb{N}^*\setminus \{1\}$, if the server draws $k$ teachers who do not all have the same vote for the term $X^k$ and its victim for the term $X$, then the class sent to the student is its victim's vote.
\end{itemize}

To address this vulnerability, we could think of an additional entity that receives the votes from the teachers and shuffles them before sending them to the server. However, the server would know if a same vote was drawn several times (remind that the drawing is with replacement), which still constitutes some information we did not account for in our DP analysis. Suppose that the server knows that all the teachers except its victim voted for a class $c$. Moreover, suppose that the offset parameter is set to $1$ and that there are $|C|$ classes in the problem. Then, there are $|C|-1$ votes different than $c$ and the victim's vote, which is unknown. Assume that the polynomial parameterization is $|C|X^2 + X$. If the $2|C| + 1$ votes that the server drew are all from different sources - teachers or dummy one-hot encodings - (remind that the server knows it) and the output class is not $c$, then the server knows with certainty that its victim did not vote for $c$ (otherwise, there would have been $|C|+1$ drawn votes for $c$ and, among the $|C|$ pairs the server drew for the term in $X^2$, no pair would have been composed of two identical votes different from $c$ and at least one pair would have been composed of two votes for $c$ and then the output would have been $c$).

These observations show that we need to constrain the server not to see the student model once it is trained. Note that the information leakage induced by the server's knowledge may not jeopardize much the data privacy in practice. We only argue here that our DP analysis does not allow us to derive DP guarantees from the point of view of the server, which might be possible with a more involved (and likely quite complex) analysis, although with probably worse guarantees.

\subsection{Extension of the threat model}
\label{sec:extension_threat_model}

We could extend the threat model and assume that the server has access to the final model by designing a more complex algorithm for which the teachers would be homomorphically selected via encrypted masks.

Another interesting idea mentioned above would be to make use of an intermediate entity that would shuffle the encrypted votes before the server receives them, with inspiration from the ESA (Encode, Shuffle, Analyze) method from \cite{bittau2017prochlo}. Nevertheless, the server would still know if it selected several times the same teacher, even without knowing which one it is, and this is still theoretically an information leakage that is not simple to analyze (cf. Section~\ref{sec:dp_not_for_server}). A way to solve this issue and to actually leverage the anonymity provided by the shuffling would be to design an algorithm that uses sampling \emph{without} replacement and to force the teachers to send a new encryption of their votes for each polynomial rounds, which would significantly increase the complexity of the protocol and its communication cost.

Aware of this weakness of our threat model compared to SPEED's one in \cite{grivet2021speed}, we let these improvements for further work.

\subsection{Computational complexity of SHIELD}

Compared with previous argmax HE computation methods, SHIELD is unique in that its complexity only linearly depends on the number of classes for the chosen machine learning problem. Indeed, the main impact of an increase in the number of classes is that the encoding space increases by the same amount (and therefore the time overhead is linear). A secondary impact is the logarithmic increase in the number of rotations needed for the computation of $\sum_{j} \pi^{(i)}_j$ as seen in Section \ref{subsec:SIMD-SHIELD-implementation}. 
All previous work uses one (or a combination) of two methods to evaluate an exact argmax over a number of values: a tournament method or a league method. We refer the reader to \cite{CZ22, IZ21} for specific implementation details. Here we focus on their complexity with respect to the number of classes.
\begin{itemize}
    \item a league is a system of comparison where every value is compared with every other value. The winner is the value that was greater than every other one. Think of a football league in Europe for this kind of system. The use of a league method yields a quadratic complexity in the number of classes. This leads to very high performance overheads as the number of classes increases. However, contrary to the tournament method, increasing the number of classes does not affect the multiplicative depth of the circuit to be evaluated. This is what makes this method useful in the homomorphic domain in spite of its complexity.
    \item a tournament is a system where values are compared two-by-two and the losers are discarded at every round. Think of the FIFA World Cup for this kind of system. Using a tournament method has a - theoretical - linear complexity in the number of classes. In practice, this is not the case. As the number of classes increases, the comparison tree used for the evaluation increases in depth logarithmically. For leveled homomorphic schemes such as BFV or BGV (those we use in this article) used in \cite{IZ21}, this means an increase in parameter size to match the multiplicative depth of the new tree. In turn, this impacts the performance of the overall scheme on top of the theoretical linear increase. After a given point, the increase in parameter size becomes prohibitive and one needs to resort to finishing the computation using a league method as they do in \cite{IZ21}.  
\end{itemize}

Compared to all other existing works therefore, ours scales much better with the number of classes and therefore fits particularly well with use-cases with high numbers of classes.

\section{Experimental results}
\label{sec:shield_experiments}

\subsection{Choice of the polynomial parameterization}
\begin{figure*}
    \centering	\includegraphics[width=\textwidth]{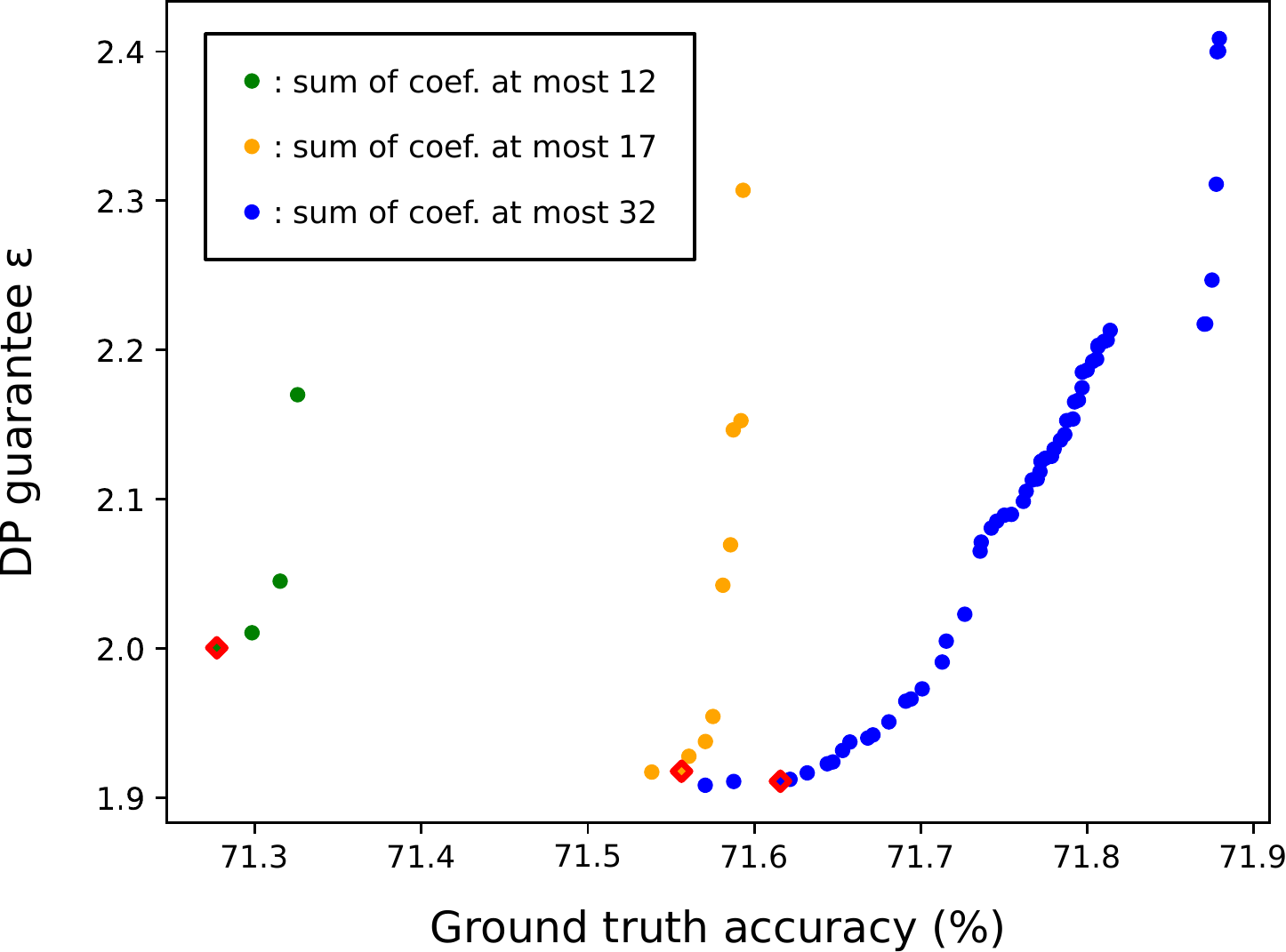}
    \caption{Pareto fronts of the polynomials for a fixed maximum sum of coefficients. The polynomials we chose for running the student model training are indicated by red-edged diamonds.}
    \label{fig:pareto_fronts}
\end{figure*}

We tested SPEED with SHIELD on MNIST dataset \cite{lecun2010mnist}. While the offset parameter has been set to $1$, a key aspect of our experiments is the choice of a polynomial parameterization that realizes a good trade-off between model accuracy, DP guarantees and computational efficiency. Since the computational time overall depends on the sum of coefficients and the degree of the polynomial parameterization, we proceeded by constraining the maximum degree and the maximum value for the sum of coefficients of the polynomials. We fixed the maximum degree to $4$ because higher degrees resulted in too high computational complexity. For several integer values ($6$, $12$, $17$, $32$), we considered all the polynomials of degree at most $4$ whose sum of coefficients is less than this value. We do not go beyond a sum of coefficients equal to $32$ to keep the computational time low. %because that allows us to batch all samples into a single ciphertext and therefore optimize the computation.
We then computed the DP guarantee $\epsilon$, $\delta = 10^{-5}$ being fixed, for each polynomial, as well as its \gta~that acts as a proxy for the student model accuracy which could not be determined in reasonable time for so many polynomials. Finally, we focused on the polynomials belonging to the Pareto front for these two criteria - DP guarantee $\epsilon$ and \gta~- and picked the ones that yielded among the best DP guarantees without harming the accuracy too much. In practice, as it can be seen on Figure~\ref{fig:pareto_fronts} the DP guarantee guided more our choice because the \gta, besides being only a heuristic for the actual student model accuracy, did not vary much among the polynomials of the Pareto front. Note that the \gta~of the exact argmax is $72.35 \%$. The chosen polynomials are respectively $2X^3+3X^2+X$, $2X^4+6X^3+3X^2+X$, $6X^4+6X^3+4X^2+X$ and $8X^4+6X^3+4X^2+X$ for a sum of coefficients of at most $6$, $12$, $17$, $32$\footnote{The chosen polynomial among the ones with a sum of coefficients at most $32$ has a sum of coefficients equal to $19$ only. This is good news for computational complexity because it allows us to batch all samples into a single ciphertext and therefore optimize the computation.}. We did not display the Pareto front for a sum of coefficients of at most $6$ because it only contains one polynomial.

Table~\ref{table:polynomial_accuracy_dp} displays the \gta, the student model accuracy and the DP guarantee $\epsilon$ for the chosen polynomial parameterizations.
The \gta~and DP guarantee are averaged on the whole set of $8000$ samples used for semi-supervised training, the DP guarantee being remultiplied by $100$, the number of actual queries to the teachers.
The student model accuracy is averaged over ten runs, each of which used a different random subset of $100$ samples as labeled samples.
The table also displays the number of correctly labeled samples (comparing to the ground truth label) out of the $8000$ samples.
The variance of the model accuracy among the runs is quite important and may explain why the accuracy surprisingly does not increase when the polynomial is better in terms of both \gta~and number of correct labels.

\begin{table}
    \centering
    \begin{tabular}{|c|c|c|c|c|}
        \hline 
        polynomial & GTA & \begin{tabular}{@{}c@{}}number of \\ correct labels \end{tabular} & \begin{tabular}{@{}c@{}}model \\ accuracy \end{tabular} & $\epsilon$ \\
        \hline\hline
        exact argmax & $72.35\%$ & $7516$ $(93.95\%)$ & $95.36\%$ & $\infty$ \\
        \hline
        $2X^3+3X^2+X$ & $70.06\%$ & $7166$ $(89.58\%)$ & $90.91\%$ & $2.39$ \\
        \hline
        $2X^4+6X^3+3X^2+X$ & $71.26\%$ & $7327$ $(91.59\%)$ & $94.66\%$ & $2$ \\
        \hline
        $6X^4+6X^3+4X^2+X$ & $71.56\%$ & $7358$ $(91.98\%)$ & $93.39\%$ & $1.92$ \\
        \hline
        $8X^4+6X^3+4X^2+X$ & $71.62\%$ & $7367$ $(92.09\%)$ & $93.15\%$ & $1.91$ \\
        \hline
    \end{tabular}
    \vspace{0.3cm}
    \caption{Accuracy and DP guarantee obtained with several polynomial parameterizations.}
    \label{table:polynomial_accuracy_dp}
\end{table}

\subsection{SIMD SHIELD with BFV}
    
    For our implementation of the SIMD SHIELD algorithm, we use the BFV cryptosystem in the openFHE library \cite{openfhe}. The parameters we choose are the following: $\log_2(q) = 540$ ; $p=65537$ ; $m=65536$ ; $N=32768$. These parameters achieve a security level of $\lambda=128$ bits with a standard deviation of $3.2$.
    Our implementation was tested on a machine with an AMD Opteron(tm) Processor 6172 using a \emph{single thread}.
    
    We achieve performances presented in Table \ref{table:packed_performance} for a set of different polynomial parameterizations. Although we tested using the MNIST data set, the performance of an HE algorithm does not depend on the underlying data \emph{by construction}. Otherwise one could infer something on the data from seeing the computation happen in the encrypted domain. For our implementation, we need to run the SHIELD algorithm over $100$ samples. In the table however, we also present computation times for the case whereby we optimize the batching space with a higher number of samples to give an idea of what computation times could be achieved by optimizing parameters further. For now, these optimizations are not yet possible in keeping with the Homomorphic Encryption Security Standard \cite{he-standard} which recommends the use of power-of-two cyclotomic polynomials. A new standard is reported to be in the works which would open applications to the secure use of non-power-of-two cyclotomic polynomials. That would allow us to optimize our parameters further.
    
    \begin{table}
        \centering
        \begin{tabular}{|c|c|c|c|}
            \hline 
            polynomial & samples & time (s) & time/sample (s) \\
            \hline\hline
            \multirow{2}{*}{$2X^3+3X^2+X$} & $100$ & $87.2$ & $0.87$ \\
            \cline{2-4}
            & $341$ & $112$ & $0.33$ \\
            \hline
        %    \multirow{2}{*}{$7X^3+4X^2+X$} & $100$ & $108$ & $1.08$ \\
         %   \cline{2-4}
         %   & $143$ & $124$ & $0.87$ \\
         %   \hline
            \multirow{2}{*}{$2X^4+6X^3+3X^2+X$} & $100$ & $123$ & $1.23$ \\
            \cline{2-4}
            & $143$ & $135$ & $0.94$ \\
            \hline
         %   $6X^4+7X^3+3X^2+X$ & $100$ & $137$ & $1.37$ \\
         %   \hline
            $6X^4+6X^3+4X^2+X$ & $100$ & $138$ & $1.38$ \\
            \hline
            $8X^4+6X^3+4X^2+X$ & $100$ & $144$ & $1.44$ \\
            \hline
            \hline
            paper & samples & time (s) & time/sample (s) \\
            \hline
            \hline
            \cite{grivet2021speed} & $100$ & $390$ & $3.9$ \\
            \hline
            \cite{grivet2021speed} + \cite{CZ22} & $100$ & $160$ & $1.6$ \\
            \hline
            \multirow{2}{*}{\cite{IZ21}$^{*}$} & $100$ & $152$ & $1.52$ \\
            \cline{2-4}
            & $5220$ & $152$ & $0.03$ \\
            \hline
        \end{tabular}
        \vspace{0.3cm}
        \caption{Performance for the SIMD implementation of SHIELD (for 10 classes) for different polynomial parameterizations compared with previous work implementing exact argmax computations. \\ 
        $*$ Times for \cite{IZ21} are presented but cannot directly compare with our results for reasons that are expanded upon below.}
        \label{table:packed_performance}
    \end{table}

    Table \ref{table:packed_performance} also compares our method with previous existing methods for \emph{exact argmax computations}. Among these methods, the one presented in \cite{grivet2021speed} as well as its later improvement in \cite{CZ22} perform worse overall for all polynomial parameterizations that we tested. It is important to note that these methods do not use batching by construction. Therefore the time per sample is fixed and does not depend on the amount of samples processed. 
    \\\cite{IZ21} on the other hand does make use of batching. In effect, by construction, they are constrained to batching sizes much higher than ours, therefore an amortized time of $0.03$s could not be obtained over $100$ samples. Times in Table \ref{table:packed_performance} for \cite{IZ21} are taken from their Table 4 because it most closely matches our use-case. However important differences remain: we report their timings for $8$ classes as it is the closest to $10$ in the Table; timings are for a minimum computation, which is less time-consuming than an argmin computation, but no times are given for an argmin in the paper.

\subsection{Bitwise SHIELD with Cingulata}

    To show the interest of the batching approach, we also implemented the basic version of SHIELD, as described in Alg. \ref{algo:multi_degree_shield}, with Cingulata crypto-compiler and its TFHE backend. 
    
    Let us remind that Cingulata, formerly known as Armadillo~\cite{cingulataV1}, is a toolchain and run-time environment (RTE) for implementing applications running over homomorphic encryption. Cingulata
    provides high-level abstractions and tools to facilitate the implementation and the execution of privacy-preserving applications expressed as Boolean circuits.

    Table \ref{table:tfhe_performance} shows the execution times of SHIELD for different polynomial parameterizations when performed in a SISD fashion with TFHE and Cingulata. The experiments were performed with a single thread  on an Intel Xeon processor with 16 GB of memory and Ubuntu 20.04 operating system. As shown in the table, the execution time of SHIELD increases with the degree of the polynomial and the sum of the polynomial coefficients. As expected, the overall performances are highly below the ones obtained when using BFV and its batching capabilities. 
    %\todo{refer to Table~\ref{table:tfhe_performance}}
    
    \begin{table}
        \centering
        \begin{tabular}{|c|c|c|c|}
            \hline 
            polynomial & samples & time (s) & time/sample (s) \\
            \hline\hline
            $2X^3+3X^2+X$ & $100$ & 495,3 & 4,95 \\
            \cline{2-4}
        %    & $287$ &  &  \\
        %    \hline
        %    $7X^3+4X^2+X$ & $100$ & 1192,32 & 11,92  \\
        %    \cline{2-4}
        %    & $143$ &  &   \\
            \hline
            $2X^4+6X^3+3X^2+X$ & $100$ & 14287,8 & 14,29 \\
            \cline{2-4}
        %    & $143$ &  & \\
        %    \hline
        %    $6X^4+7X^3+3X^2+X$ & $100$ & 21158,7  &  21,15 \\
            \hline
            $6X^4+6X^3+4X^2+X$ & $100$ & 20936,7  & 20,93 \\
            \hline
        \end{tabular}
        \vspace{0.3cm}
        \caption{Performance for the Cingulata with TFHE implementation of SHIELD}
        \label{table:tfhe_performance}
    \end{table}

\section{Conclusion and perspectives}
We proposed SHIELD, a homomorphic stochastic operator whose lightweight design necessary for fast homomorphic computations yields DP as a natural ``by-product''.
This work reconciliates two complementary but usually independent - or even mutually constraining - privacy tools in an all-in-one operator whose inaccuracy is a crucial feature.

We hope this work will encourage new works on the design of private algorithms where FHE (or other cryptographic primitives) and DP leverage the advantages of each other. For instance, developing algorithms that would be useful in other settings than an election and broaden the scope of machine learning applications seems promising. In this perspective, an argmax algorithm that takes an histogram of the votes as input rather than the ``physical'' votes represented as vectors would have a more general applicability.

Testing SHIELD on more difficult datasets and especially datasets with numerous classes could reveal its full potential.
Besides, a more thorough theoretical study to get results that may lead us through the choice of the parameters (polynomial, offset) is desirable.
Other versions including sampling without replacement (Section~\ref{sec:extension_threat_model}) or an exponential version of SHIELD (Section~\ref{sec:exponential_argmax}) would also deserve theoretical and experimental analyses.
Studying SHIELD in terms of strategy-proofness and fairness could be interesting too and would extend the added value of SHIELD to the area of computational social choice and voting rules.

\bibliographystyle{plain}
\bibliography{./bibliography/biblioDP, ./bibliography/biblioCrypto, ./bibliography/biblioCrypto2}

\clearpage

\begin{appendices}
\section{On counter-productive noise for data-dependent differential privacy guarantees}

\textbf{Null data-dependent privacy cost of the exact argmax:} While doing experiments on a subset of MNIST with polynomial parameterizations that yield better and better accuracies (up to the probability of getting the true argmax being more than 99,99\%) we remarked that the value epsilon of the privacy cost did not increase much and did not seem to approach infinity. This surprising result suggested that the exact argmax operator had a finite privacy cost. Actually, on the subset we were working on, for every sample, the dominant class had at least two more votes than the second dominant class. We will say in the following that the distribution has a \emph{highly dominant} argmax. This implies that, any database which is adjacent (i.e. differs from at most one teacher) to the database $d$ we were working on has the same dominant class as $d$ for every sample. As a consequence, the output of the argmax does not leak \emph{any} information about which of two adjacent databases was used as input. In other words, the privacy cost of the exact argmax operator is null in this case.

\textbf{Counter-productivity of the noise regarding privacy:} On the contrary, the so-called private argmax operator (noised by an additional random noise as in PATE \cite{papernot2016semi, papernot2018scalable} and SPEED~\cite{grivet2021speed} or intrinsically stochastic as in SHIELD) may output any class and the probabilities of outputting a class depends on the frequencies of the votes for all classes. As a consequence, even changing only one teacher will change the probabilities of outputting some (or rather all) of the classes, even if the effect is mild. Therefore, the output of the DP argmax operator does give information on the probability of outputting a class and then on the frequencies of the classes in the votes. We end up in a (particular) situation where applying noise is counter-productive in the sense that it increases the privacy cost of revealing the output (by an infinite factor actually). Note, however, that this was not the case for the entire MNIST training set but only for a certain subset of it.

\textbf{The case of data-independent DP guarantees:}
This consideration only applies to \emph{data-dependent} DP guarantees. In the data-independent case, the privacy cost of the exact argmax would be infinite because we would consider the maximum over all the pairs of adjacent databases i.e. all the possible pairs of distributions of $n$ votes among $K$ classes that differ by one vote on two classes. In this perspective, the question of the definition domain of the databases is crucial. Only giving data-dependent DP guarantees for the aforementioned subset of MNIST dataset, where, for every sample, the vote distribution has a highly dominant argmax, amounts to give a data-independent DP guarantee with a definition domain of the databases included in the set of databases such that the vote distributions have a highly dominant argmax. This is obviously restricting the problem to a too easy subset of situations, and, as we showed above, this restricted problem is trivially solved by the deterministic exact argmax.

\textbf{Example of the age's sign:}
The noise addition degrading privacy guarantees is very counter-intuitive and may surprise \textit{a priori}. Let us take a simple example to understand how the noise affects privacy. Revealing the sign of the age of a person is infinitely private (epsilon and delta null) if we assume that the adversary already knows that a person must have a positive age (quite natural assumption!). Imagine now that we noise the age with a unimodal noise, whose mode is zero, say a Gaussian noise, before computing the sign. The lesser the unnoised age, the more likely the sign of the noised age will be negative. This implies that revealing the sign of the noised age does leak some information about the unnoised age. Clearly, the noise addition does harm the privacy guarantees in this case.
Nevertheless, note that this does not contradict the post-processing immunity of DP. Indeed, the noise is not added at the end, over the infinitely private sign of the age, rather, it is added before the computation of the sign, inside the mechanism and not afterwards. Thus, the noise addition cannot be considered as a post-processing.
\end{appendices}

\end{document}